\documentclass{aastex631}
\usepackage{graphicx}
\usepackage{xspace} 
\usepackage{enumerate}
\usepackage{multirow}

\def \lp{\>\> .}
\def \lc{\>\> ,}

\newcommand{\degree}{\mbox{$^{\circ}$}}
\newcommand{\am}{\mbox{\arcmin}}
\newcommand{\as}{\mbox{\arcsec}}

\newcommand{\kms}{\mbox{\,km\,s$^{-1}$}}

\newcommand\cmv{\mbox{cm$^{-3}$}}
\newcommand\cmc{\mbox{cm$^{-2}$}}

\newcommand{\ps}{s$^{-1}$}
\newcommand{\psr}{sr$^{-1}$}
\newcommand{\pHz}{Hz$^{-1}$}
\newcommand{\MJysr}{MJysr$^{-1}$}

\newcommand{\co}{$^{12}$CO}
\newcommand{\coo}{$^{13}$CO}

\def \18OI{[$^{18}$O\,{\sc i}]}
\def \17OI{[$^{18}$O\,{\sc i}]}
\newcommand{\hh}{\mbox{{\rm H}$_2$}}

\def \13CII{[$^{13}$C\,{\sc ii}]}

\newcommand{\Inu}{$I_{\nu}$}

\def \lp{\>\> .}
\def \lc{\>\> ,}
\def \beq{\begin{equation}}
\def \eeq{\end{equation}}
\begin{document}

\title{The \hh\ Glow of a Quiescent Molecular Cloud Observed with JWST}
\shorttitle{JWST Observations of \hh\ in Taurus}

\author[0000-0002-6622-8396]{Paul F. Goldsmith}
\affiliation {Jet Propulsion Laboratory, California Institute of Technology, 4800 Oak Grove Drive, Pasadena, CA 91109, USA}

\author[0009-0007-6655-366X]{Shengzhe Wang}
\affiliation{School of Astronomy and Space Science, University of Chinese Academy of Sciences (UCAS), Beijing 100049, China}
\affiliation{National Astronomical Observatories, Chinese Academy of Sciences, Beijing 100101, China}

\author[0000-0002-9373-3865]{Xin Wang}
\affiliation{School of Astronomy and Space Science, University of Chinese Academy of Sciences (UCAS), Beijing 100049, China}
\affiliation{National Astronomical Observatories, Chinese Academy of Sciences, Beijing 100101, China}
\affiliation{Institute for Frontiers in Astronomy and Astrophysics, Beijing Normal University, Beijing 102206, China}

\author[0000-0003-2337-0277]{Raphael Skalidis}
\thanks{Hubble Fellow}
\affiliation {Jet Propulsion Laboratory, California Institute of Technology, 4800 Oak Grove Drive, Pasadena, CA 91109, USA}
\affiliation{TAPIR, Mailcode 350-17, California Institute of Technology, Pasadena, CA 91125, USA }

\author[0000-0001-8509-1818]{Gary A. Fuller}
\affiliation{Jodrell Bank Centre for Astrophysics, Department of Physics and Astronomy, The University of Manchester, Oxford Road, Manchester M13 9PL, UK}
\affiliation{I. Physikalisches Institut, University of Cologne, Z\"ulpicher Str. 77, 50937 K\"oln, Germany}

\author[0000-0003-3010-7661]{Di Li}
\affiliation{Department of Astronomy, Tsinghua University, Beijing 100084, People's Republic of China}
\affiliation{ National Astronomical Observatories, Chinese Academy of Sciences, Beijing 100101, People's Republic of China}
\affiliation{ Research Center for Astronomical Computing, Zhejiang Laboratory, Hangzhou 311100, People's Republic of China}
\affiliation{New Cornerstone Science Laboratory, Shenzhen 518054, People's Republic of China }

\author[0000-0002-9390-9672]{Chao-Weo Tsai}
\affiliation{National Astronomical Observatories, Chinese Academy of Sciences, Beijing 100101, China}
\affiliation{Institute for Frontiers in Astronomy and Astrophysics, Beijing Normal University, Beijing 102206, China}
\affiliation{School of Astronomy and Space Science, University of Chinese Academy of Sciences (UCAS), Beijing 100049, China}

\author[0000-0002-6540-7042]{Lile Wang}
\affiliation{The Kavli Institute for Astronomy and Astrophysics, Peking University, Beijing 100871, People's Republic of China}
\affiliation{Department of Astronomy, School of Physics, Peking University, Beijing 10087, People's Republic of China}

\author[0000-0003-4811-2581]{Donghui Quan}
\affiliation{Zhejiang Laboratory, Hangzhou 311121, People's Republic of China}

\correspondingauthor{Paul Goldsmith, Xin Wang}
\email{paul.f.goldsmith@jpl.nasa.gov, xwang@ucas.ac.cn}
\begin{abstract}
We report JWST MIRI/MRS observations of the \hh\ $S(1)$ 17.04 $\mu$m transition in two regions in the boundary of the Taurus Molecular Cloud.
The two regions, denoted ``Edge'' (near the  relatively sharp boundary of the \coo\ $J = 1\rightarrow 0$ emission) and ``Peak'' (the location of the strongest \hh\ emission observed with {\it Spitzer}), have average intensities of 14.5 MJy~\psr\ and 32.1 MJy~\psr, respectively.
We find small scale structures of characteristic size 1.0\as--2.5\as, corresponding to 140 AU--350 AU, with characteristic intensity above the extended background of 10 MJy~\psr, corresponding to a $J$ = 3 column density of 1.6$\times$10$^{17}$ \cmc.  
The most plausible explanation for the observed intensities from level 845 K above the $J$ = 1 ortho--\hh\ ground state level is excitation by collisions with \hh\ molecules (the hydrogen in this region being predominantly molecular).
Two mechanisms, turbulent dissipation and shocks, have been proposed for heating of localized regions of the ISM to temperatures $\simeq$1000 K to explain abundances of and emission from particular molecules. 
While we cannot determine unique values of density and kinetic temperature, the solutions in best agreement with predictions of shock models are $n$(\hh) = 370 \cmv\ and $T_{kin}$ = 1000 K.  
The total \hh\ column density of the small--scale structures under these conditions is $\simeq$8$\times$10$^{17}$ \cmc.
This first direct detection of tiny scale structure in the quiescent molecular interstellar medium has significant implications for the physical structure of this phase of the ISM and the maintaining of supersonic motions within it.

\keywords{Interstellar medium (847); shocks; Interstellar line emission (844); Molecular
clouds (1072)}

\end{abstract}
\section{Introduction}
Understanding the small--scale structure of the interstellar medium (ISM) is fundamental for many reasons.  
On scales smaller than 100 AU, different physical processes may reveal themselves, including the dissipation of turbulence \citep{Falgarone07, Godard09, Godard14} and the fragmentation of gas \citep[][and references therein]{beuther19} during the process of star formation.
Very small---scale structure has been seen in atomic and ionized phases of the ISM, as discussed in the reviews by \citet{Heiles07} and \citet{Stanimirovic18}.  
For atomic gas, these are referred to as Tiny Scale Atomic Structure (TSAS) and for ionized gas, Tiny Scale Ionized Structure.  

Radio pulsars are a convenient background source for observations of optical depth variations of 21cm absorption, and combined with muti--epoch observations and their proper motion allow probing variations on quite small angular scales \citep{Frail94} but see \citet{Stanimirovic04,Stanimirovic10}.  
Variations have recently been observed down to scales $\le$ 20 pc \citep{Liu21}.
VLBI observations of extended background radio sources allow direct imaging of variations in the 21 cm signal, which are felt likely to be due to opacity variations on a scale of 25 AU \citet{Brogan07}.
Fluctuations in the column density of ionized gas can produce lensing that changes the observed flux of distant ratio sources and changes in pulsar dispersion measure characterized as Extreme Scattering Events (ESE) suggest structure on a scale of 10's to 100's of AU \citep{Stanimirovic18}. 

Much more limited studies of the Tiny Scale Molecular Structure (TSMS) have been carried out.
\citet{Moore93} and \citet{Moore95} have reported time variations in the 6 cm H$_2$CO absorption profile against two radio sources that indicate structure in the molecular ISM on the order of 10 AU in size.
Optical and UV absorption lines \citep{savage77,wakker06, Stanimirovic18} probe extremely fine lines of sight, and while they do not as readily lend themselves to studying variations and structures in the absorbing gas, people have looked for variability in absorbing features \citep[e.g.][]{Farhang23}.  

The nature of the material producing the observed very small--scale variations in the diffuse ISM remains controversial.
It is possible rather than being discrete structures of a characteristic size these are the tail of a distribution of turbulent variations \citep{Deshpande07}.  
The relationship of very small-scale structure in the atomic and ionized ISM to that in the denser, cooler molecular ISM remains unclear, with imaging of the latter being so limited.  
This situation has begun to change in the past few years.

Molecular hydrogen is the dominant constituent of the molecular ISM. 
Ground-based facilities have allowed observations with  subarcsecond angular resolution of \hh\ rotational and rovibrational emission from shocks \citep{Santangelo14}, jets \citep{Lorenzetti03}, and small--scale clumps \citep{Vannier01, Lacombe04} in the central region of Orion OMC1.
Velocity--resolved \hh\ has been observed towards shocked regions \citep{Neufeld24}.
\citet {Gry02} and \citet{Falgarone05} have observed \hh\ emission along relatively long paths through the Galactic ISM, suggesting that this may be relatively common in the Milky Way.

\citet{Goldsmith10} observed the nearby (D$\simeq$140 pc\footnote{The Taurus cloud has a significant line of sight extent, estimated for the main region of the cloud to be 130 pc to 172 pc by \citet{Galli19} and 131 to 168 pc by \citet{Zucker21} .  The largest values are for relatively limited, isolated regions.  The distance of Heiles Cloud 2, relatively close to the positions we have observed is between 138 and 142 pc \citep{Galli19}.  We adopt the 140 pc value as likely appropriate for the boundary region we have observed and  for consistency with previous studies.}) Taurus Molecular Cloud;  using {\it Spitzer} IRS with a large extraction aperture several transitions were detected, but the low angular resolution precluded being able to draw any conclusions about the size of the structures producing the hot \hh\ observed.
The Taurus molecular cloud can reasonably be characterized as quiescent, in that among the $\sim$ 400 stars accepted as being in the region defined by CO emission and dust extinction, there are 6 G type stars and 2 B type stars, with the remaining stars being later--type \citep{Guedel07}.
This means that there is little stellar UV produced \citep{Xia22}, and gas temperatures throughout the cloud are low, $\le$ 15 K \citep{Goldsmith10} as are dust temperatures (12 K -- 15 K; \citet{Flagey09}).

Molecular outflows, associated with early phases of protostars, are rare in Taurus, with only 20 known \citep{Narayanan12}.
These may be capable of sustaining the observe supersonic (turbulent) linewidths in Taurus (Section \ref{shocks} and Appendix A), but do not produce multi--\kms\ perturbations of the velocity field of the cloud.
There does not appear to be evidence of cloud--scale expansion or contraction although there is possible rotation, based on the kinematics of stars associated with the Taurus molecular cloud \citep{Galli19}.

Observations of very small--scale structure analogous to those seen in the atomic ISM but in the cooler and denser molecular ISM, and which are predicated theoretically to be sites of turbulent dissipation \citep[e.g.][]{Godard09}, are critical for understanding the energetics of these regions in which star formation occurs, but have to date been elusive.
The situation has now changed dramatically with the availability of the MIRI MRS instrument \citep{Wright23} on the JWST \citep{Gardner23}.
For example, pure rotational transitions of \hh\ from $S(1)$ through $S(8)$ have been observed in a protoplanetary disk by \citet{Berne23}.
At the 17.035 $\mu$m wavelength of the $S(1)$ pure rotational line of \hh, JWST/MIRI offers 0.7\as\ angular resolution, corresponding to a spatial resolution of $\simeq$ 100 AU or 1.5$\times$10$^{15}$ cm at the 140 pc distance of the Taurus Molecular Cloud, along with a spectral resolving power $R$ equal to 2500. 
We have used these capabilities of JWST to study the \hh\ $S(1)$ emission from two regions in the boundary of the Taurus Molecular Cloud that had previously been observed with much lower angular and spectral resolution using {\it Spitzer}  \citep{Goldsmith10}. 

Our goal is to study the  properties of \hh\ emission at scales where energy dissipation occurs. 
Such regions representative of the general ISM  were previously inaccessible for this type of investigation, but now, thanks to the high angular resolution and sensitivity of JWST, are possible.
We observed only a single transition, but have resolved the small-scale (dissipation) fluctuations of the \hh\ emission.
While nominally lying in the quiescent boundary of the molecular cloud, the fact that the \hh\ $J$ = 3 level lies $\simeq$845 K above the ortho--\hh\ ground state level demands significantly different excitation than that characterizing the 10 K -- 50 K CO \citep{Goldsmith08} seen in the same direction.

This paper is organized as follows.  
In $\S$\ref{data} we discuss the JWST data taking and processing, although details are given in the Appendix.  
In $\S$\ref{coldens} we discuss the procedure we have used to derive the column density of \hh\ emission.
In $\S$\ref{extotal}  we present a brief analysis of different mechanisms that may be responsible for excitation of the $S(1)$ line and the implications for the total column density of excited \hh\ molecules.
In $\S$\ref{highT} we discuss possible sources for the high temperatures that are required for collisional excitation of \hh.
In $\S$\ref{summary} we summarize and discuss our results.
In the Appendices we give details on the MIRI/MRS data processing and on the \co\ and \coo\ emission seen in the same direction as the observed \hh.

\section{JWST Data}
\label{data}

\begin{figure*}[htb!]
\center
\includegraphics [angle=270, width=12 cm]{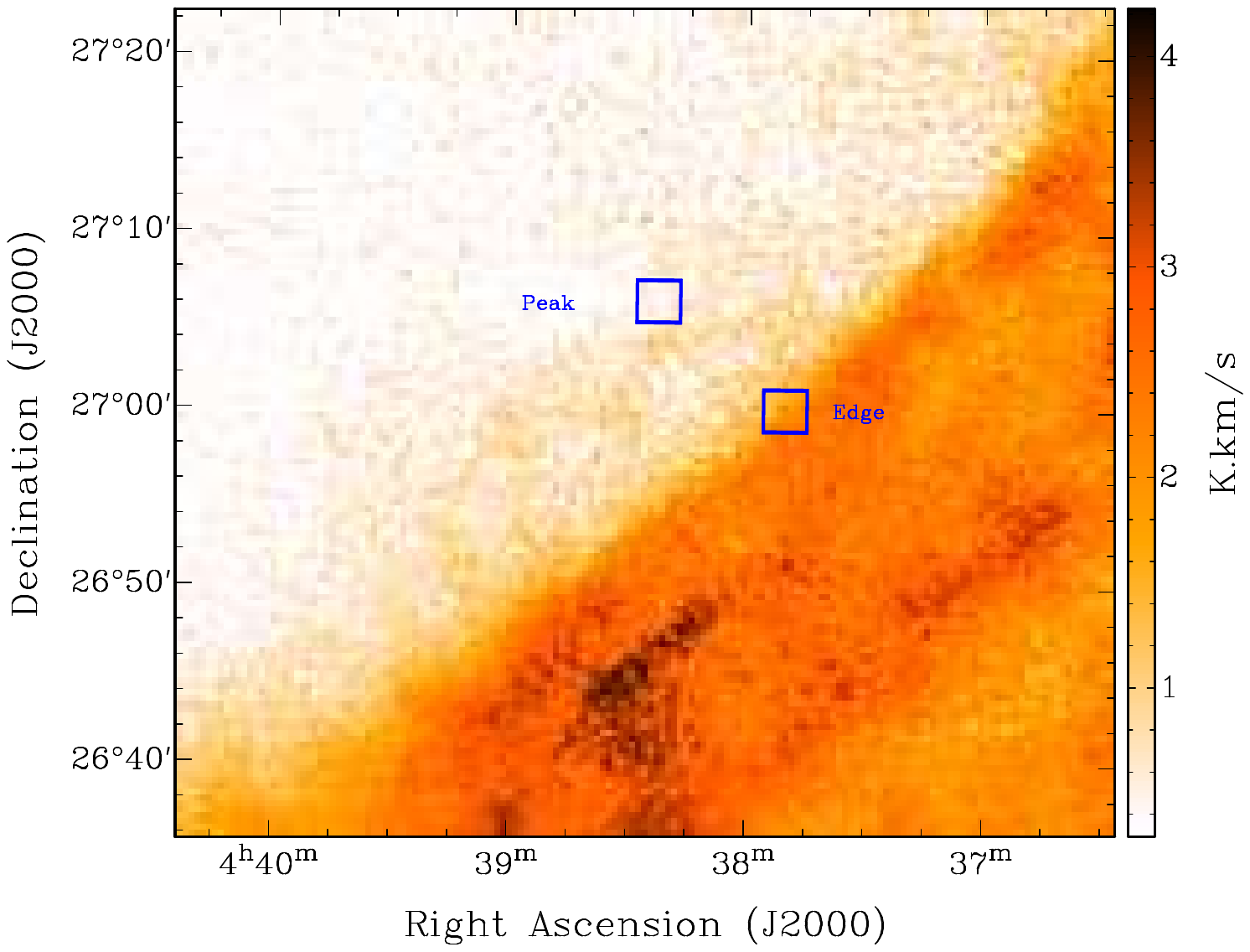}
\caption{\label{fig_regions} The two regions observed in the \hh\ $S(1)$ transition in the  boundary of the Taurus Molecular Cloud overlaid on an image of the \coo\ $J$ = 1$\rightarrow$0 integrated intensity \citep[adapted from][]{Goldsmith08}.  The size of the regions, denoted ``Edge'' and ``Peak'' has been significantly expanded to make them visible.}
\end{figure*}

The data discussed here were acquired by the JWST Cycle-2 Program GO-3050 (PI Goldsmith).
Two regions at the straight edge boundary seen in \coo\ were observed; these had been previously observed by \citet{Goldsmith10} using {\it Spitzer}.  
Figure \ref{fig_regions} shows the two regions observed overlaid on an image of the \coo\ $J$ = 1$\rightarrow$0 integrated intensity. 
The ``Edge'' position is approximately at the position of marked drop in the integrated intensity of \coo\ emission.
The ``Peak'' position includes a region at the distance from the cloud edge at which the strongest $S(0)$ emission was found with {\it Spitzer}.  
Table \ref{coords} gives the coordinates of the regions observed.

\begin{deluxetable}	{lcc}[htb!]
														
\tablecaption{Sources Observed in the Taurus Boundary Region \label{coords}}											
\tablehead{
\colhead{Source}   & \colhead{$\alpha$(2000)}    &  \colhead{$\delta$(2000)} \\
\colhead{            }   & \colhead{(h~~ m ~~s)}        &  \colhead{(\degree~~~ \arcmin~~~ \arcsec~)}      				
}
\startdata
Peak	&  04 37 50.40	& +27 00 02.5\\
Edge	&  04 38 22.75	& +27 06 12.0\\
\enddata
\end{deluxetable}

The MIRI Medium Resolution Spectrometer (MRS) consists of four channels that span the range 4.9 to $ 27.9~\mu\text{m} $.
The wavelength range of MRS covers four emission lines detectable in the {\it Spitzer} observations: the $S(0)$ line at $28.22~\mu\text{m}$ (sub-band C, Channel 4), the $S(1)$ line at $ 17.04~\mu\text{m} $ (sub-band C, Channel 3), the $S(2)$ line at $ 12.2~\mu\text{m}$ (sub-band A, Channel 3), and the $S(3)$ line at $ 9.67~\mu\text{m}$ (sub-band B, Channel 2). 
Considering that the $S(0)$ line is at the far red edge of Channel 4 sub-band C, and the $S(3)$ line is much weaker (a factor of a few hundred weaker than $S(1)$), our focus was primarily on the detection of the $S(1)$ and $S(2)$ lines. 
These lines are distributed across Channel 3 sub-band C (with a wavelength range of $ 15.41-17.98~\mu\text{m} $) and Channel 3 sub-band A (with a wavelength range of $11.55-13.47~\mu\text{m}$).

The FoV of MRS Channel 3 is $5.2\arcsec\times6.2\arcsec$, whereas the previous Spitzer IRS extraction aperture covers $10.7 \arcsec\times108\arcsec$. 
To mitigate the giant leap in angular resolution from the {\it Spitzer} IRS observations to the MRS imaging spectroscopy, we designed a mosaic consisting of 4 by 4 individual MRS footprints for each sub-band C visit. 
As a result, our FoV is expanded to $17.7 \arcsec\times 20.4\arcsec$ for the primary MRS observations in each region. 
We utilized a single MRS pointing (with FoV $5.2\arcsec\times 6.2\arcsec$) to perform deep sub-band A exposures necessary to detect the much weaker $S(2)$ line.
 
Observations were conducted for both the Peak and Edge regions, targeting the $S(1)$ and $S(2)$ transitions. 
For $S(1)$, the mosaicing was combined with a 4-point dither pattern. 
Each exposure lasted 510 seconds, leading to a total exposure time of 8160 seconds. 
Additionally, background observations were performed for $S(1)$ with an identical exposure time and using the same 4-point dither pattern. 
The background observation region was selected based on prior {\it Spitzer} results. 

For $S(2)$, observations were split into two separate exposures, each using the 4-point dither pattern with an exposure duration of 7549 seconds, resulting in a total exposure time of 15,098 seconds.
 During the observations, an anomaly occurred in the S(2) observation of the Peak region. 
 Consequently, we only obtained $S(2)$ results for the Edge region, and these will be discussed in a separate paper.

A 10\% overlap between adjacent pointings was recommended by the STScI program scientist to ensure contiguous coverage for mosaicing.
As discussed in Appendix A, the actual overlap for the Band 3 spectra that included the 17.07 $\mu\text{m}$ $S(1)$ line was much greater, in excess of 60 \%.  This resulted in a slightly non-uniform sensitivity across the mosaiced area.
This worked very well, with minimal obvious artifacts within all but the edges of the mosaic.
In the portions of the mosaic within 3 pixels of the edge there was a sufficient number of pixels with obviously erroneously high values that we ended up removing edge strips from the final image.  
This may be due in part to the smaller integration time at the edges of the mosaic, as discussed in Appendix A.

\begin{figure*}[htb!]
\center
\includegraphics [angle=0, width=14 cm]{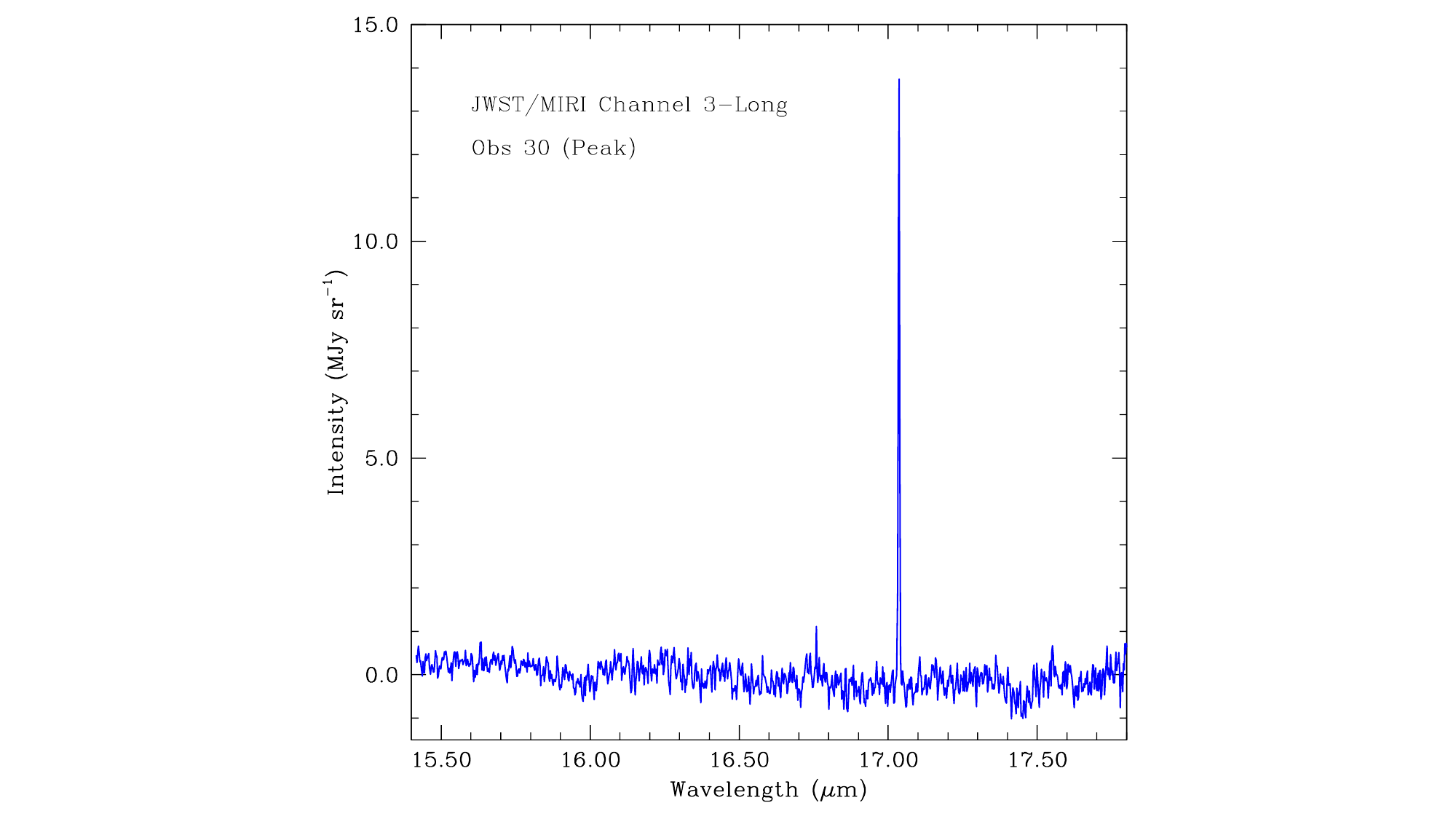}
\caption{\label{total spectrum} Output covering essentially entire spectral range of  JWST/MIRI Channel 3--Long from the Peak Region of the Taurus Molecular Cloud integrated over the area of the mosaiced observation.  Only a linear baseline has been removed.  The \hh\ $S(1)$ emission  is the only strong line that is visible. }
\end{figure*}

\begin{figure*}[htb!]
\center
\includegraphics [angle=0, width=14 cm]{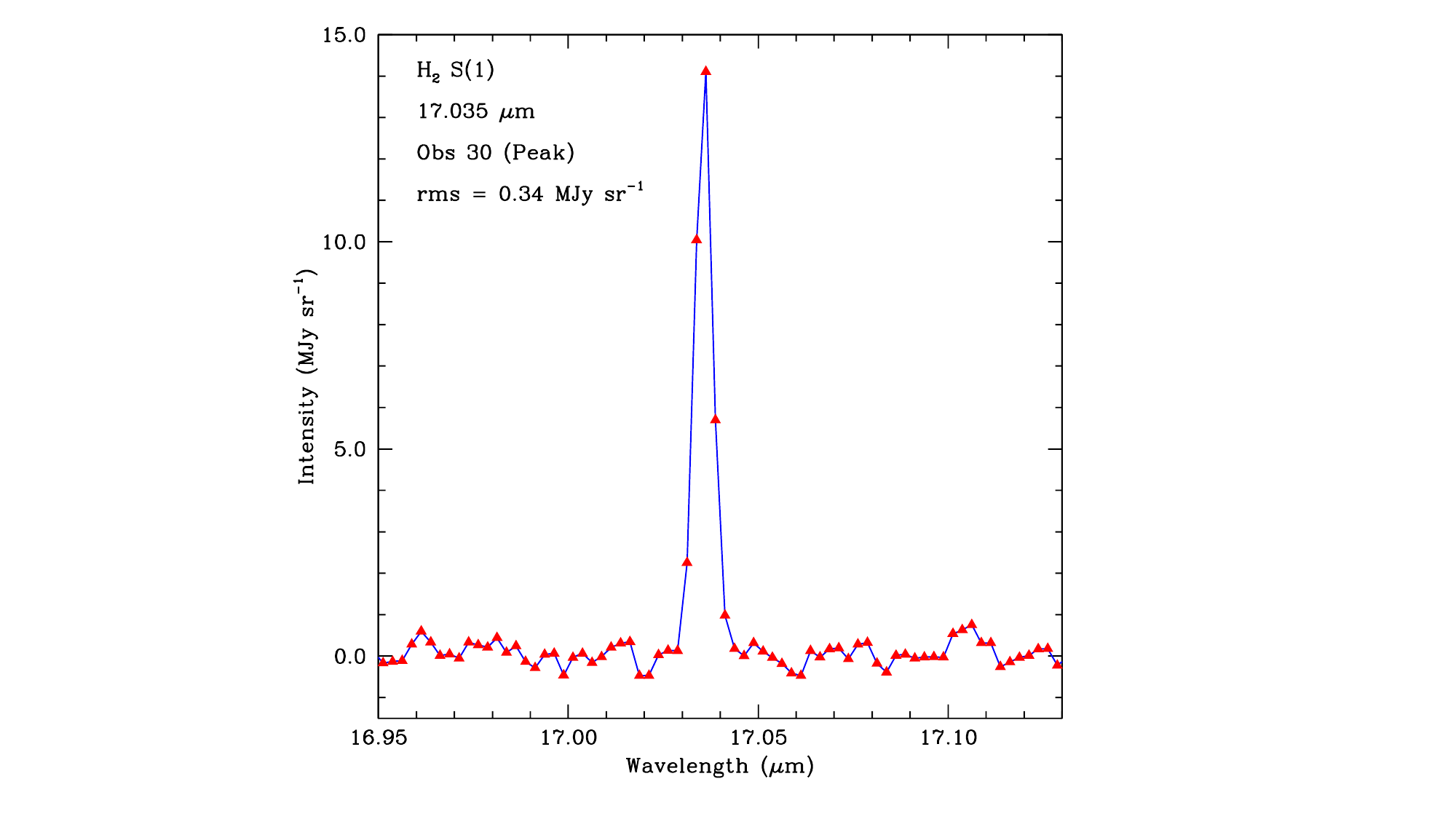}
\caption{\label{H2_17_spec_exp} Expanded view of emission around the wavelength of the \hh\ 17 $\mu m$ $S(1)$ line from the Peak Region of the Taurus Molecular Cloud integrated over the area of the mosaiced observation.  Only a linear baseline has been removed.  Note that for a resolving power $R$ equivalent to a velocity resolution of 120 \kms\  the observed spectrum is essentially the response of the MIRI spectrometer rather than being the emission line profile.}
\end{figure*}

The spectral baselines were excellent due to the subtraction of nearby reference position.
This is seen in Figure \ref{total spectrum}, which covers essentially the entire JWST/MIRI Channel 3--Long output.
The region around the $S(1)$ line is not confused by any other spectral lines, and the \hh\ line is readily apparent.
The baseline on both sides of the  $S(1)$ line was very flat so that only a linear spectral baseline was required to completely remove any continuum signal.
This was done by fitting a 1$^{st}$ order polynomial to 5 channels on each side of the $S(1)$ line starting 5 channels from the channel closest to the rest wavelength of the line.
An expanded view of the region around the $S(1)$ line is shown in Figure \ref{H2_17_spec_exp}, again illustrating the very high quality of the spectral baseline.
To obtain the total intensity in the $S(1)$ line, we summed the data from five channels centered on that closest to the wavelength of the $S(1)$ line.
Additional information on data processing is given in Appendix A.
The resulting images are shown in Figures \ref{fig_obs20} and \ref{fig_obs30}.
%
%
\begin{figure*}[htb!]
\center
\includegraphics [angle=0, width=16 cm]{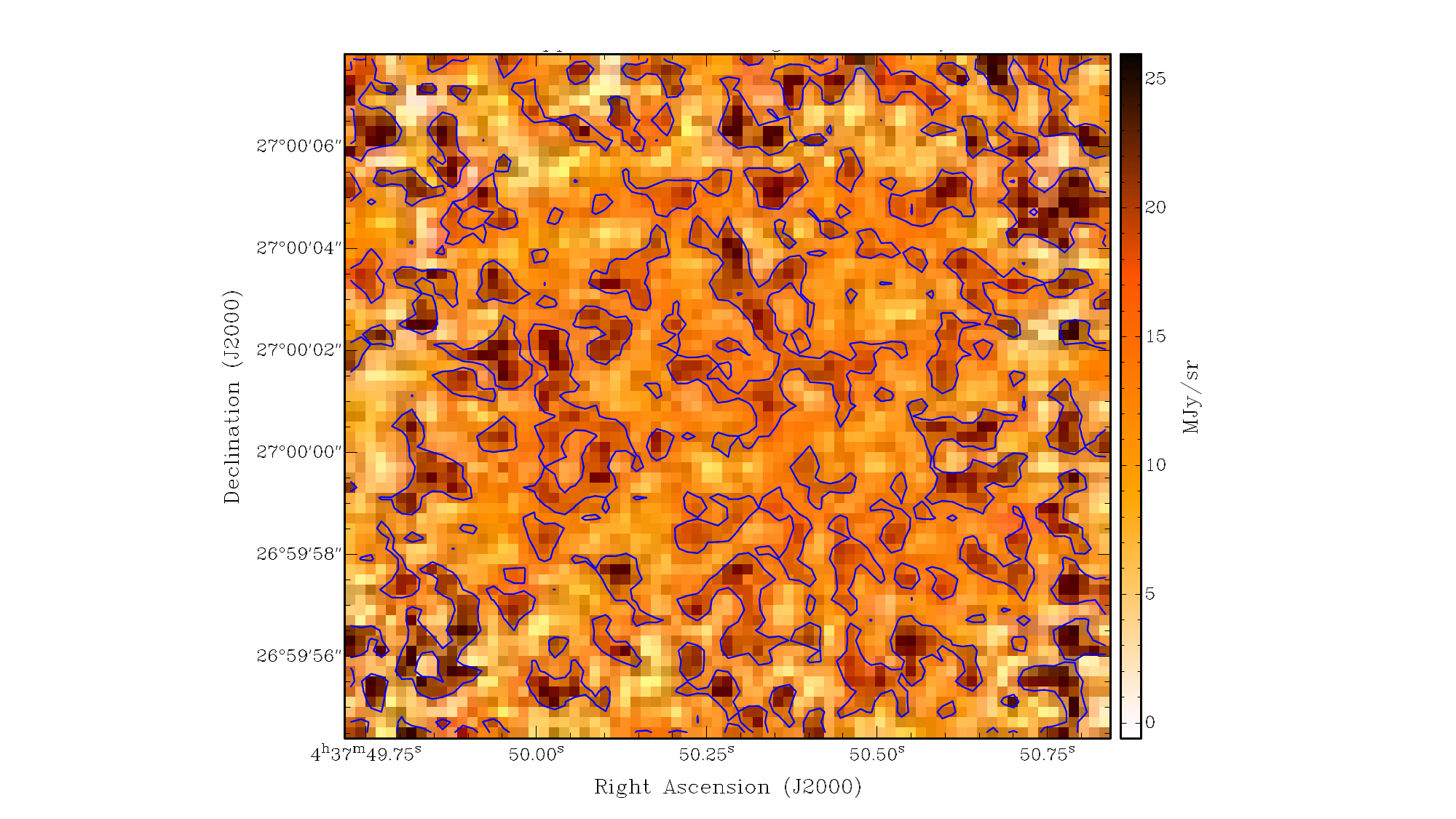}
\caption{\label{fig_obs20} \hh\ $S(1)$ emission from Edge Region of the Taurus Molecular Cloud.
The contours are at a level of 20 MJy~\psr}
\end{figure*} 

\begin{figure*}[htb!]
\center
\includegraphics [angle=270, width=14 cm]{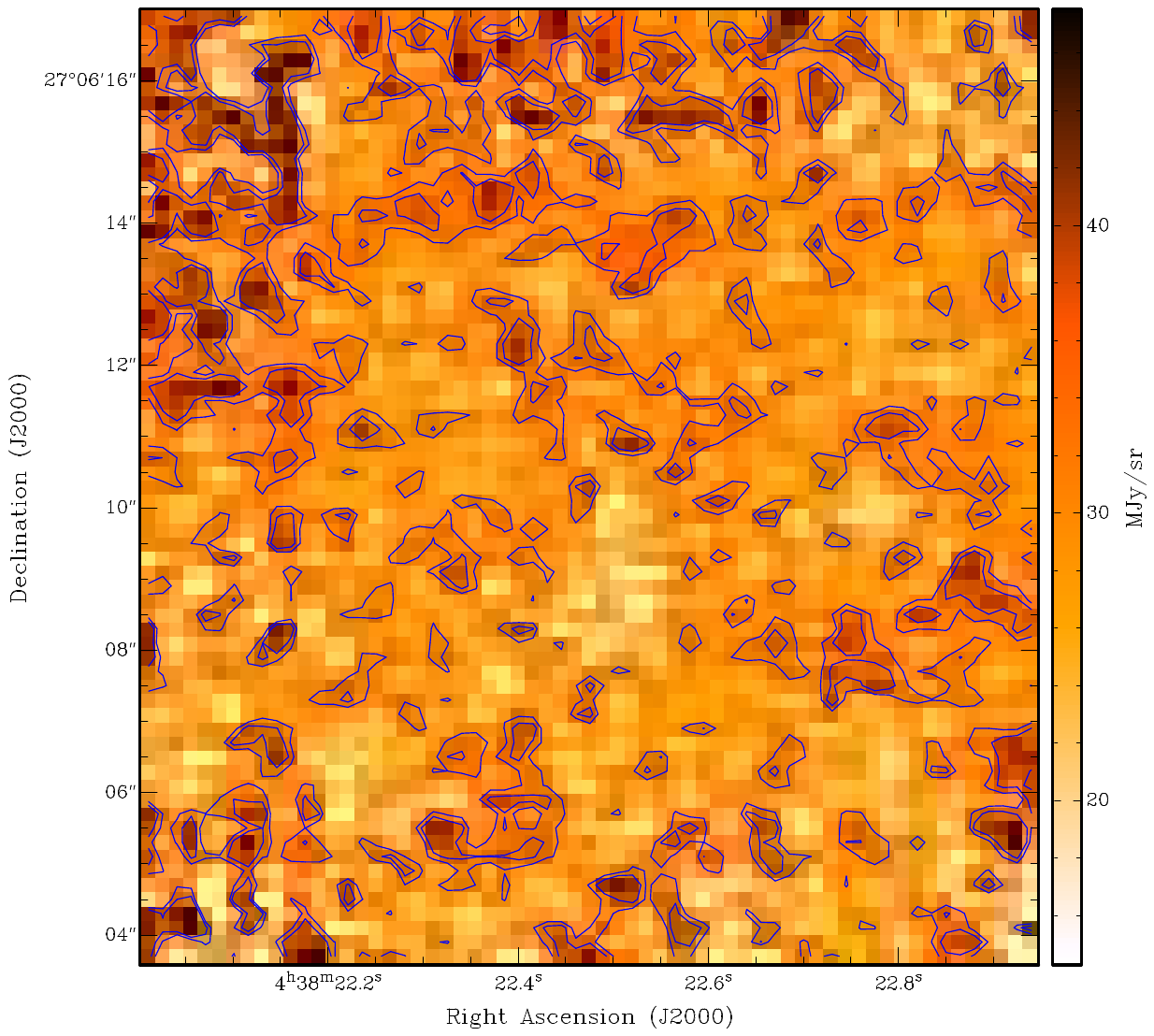}
\caption{\label{fig_obs30} \hh\ $S(1)$ emission from Peak Region of the Taurus Molecular Cloud.
The contours are at levels of 30 and 36 MJy~\psr.}
\end{figure*}

The sensitivity of the JWST/MIRI instrument and the exposure time of over 4 hours per region resulted in very significant detection of the $S(1)$ line throughout most of the FOV covering both Edge and Peak regions, although the latter is significantly stronger.

\begin{figure*}[htb!]
\center
\includegraphics [angle=0, width=14 cm]{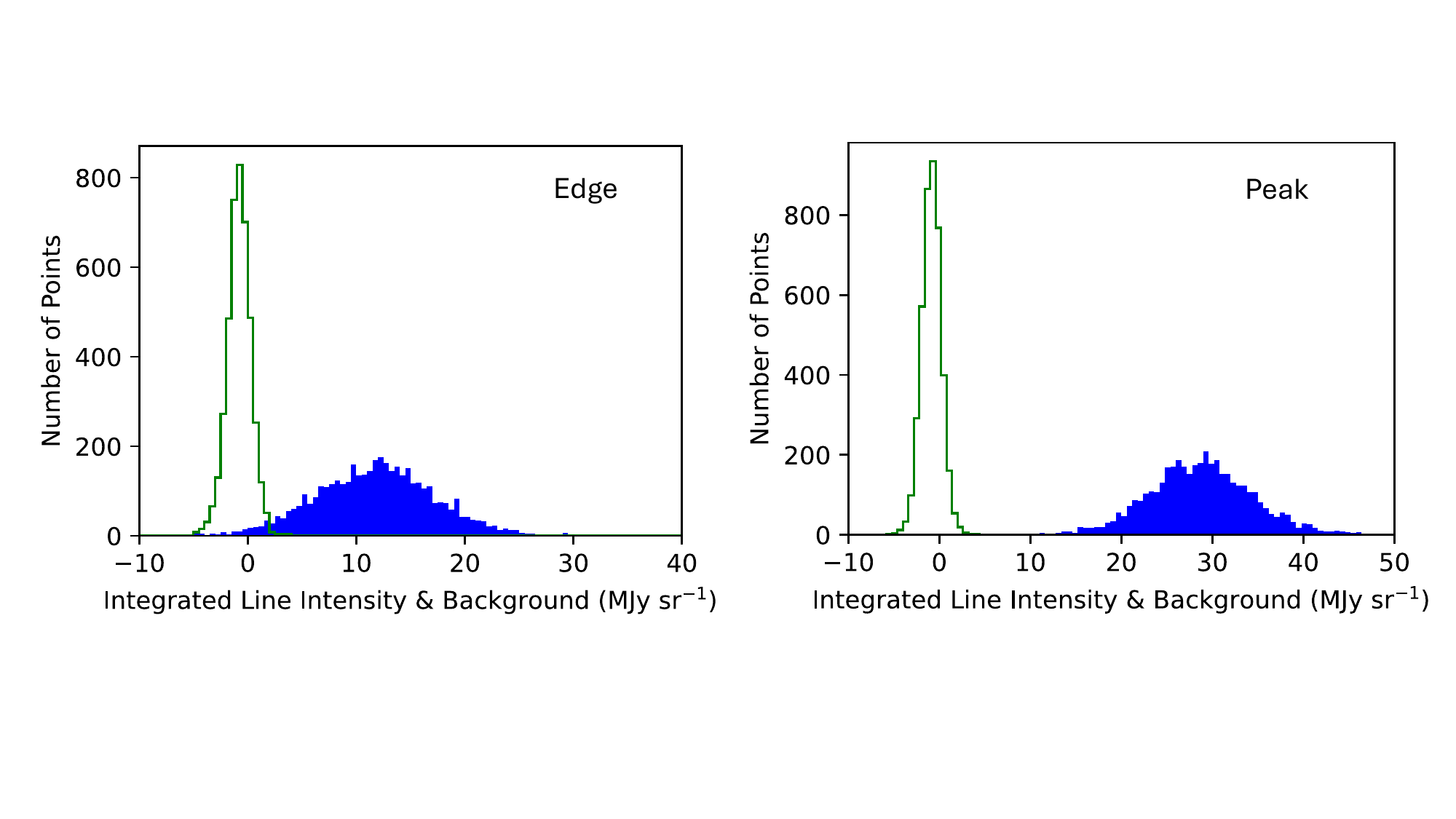}
\caption{\label{Inten_Stats} Distribution of intensity from off--line (``background'') channels (green) and summed on--line (``line'') channels (blue) for the Edge (left) and Peak (right) regions in the Taurus Molecular Cloud boundary.}
\end{figure*} 

Figure \ref{Inten_Stats} shows the distribution of intensities for the sum of the five channels ``on'' the line (line; blue) and 10 channels ``off'' the line (background; green).
The statistics of the observations are summarized in Table \ref{stats}.


\begin{deluxetable}	{lrr}[htb!]
\center														
\tablecaption{Statistics of Observations of the $S(1)$ \hh\ line in the Taurus Boundary\label{stats}}											
\tablehead{
\colhead{Parameter\tablenotemark{a}}   	& \colhead{Edge}    	&  \colhead{Peak} \\		
                                                                	& \colhead{Region} 	& \colhead{Region}\\	
}
\startdata
$<$background$>$               			& -0.8   			& -0.9 \\
$\sigma$(background)          			& 1.1     			&  1.1\\
min(background)                   			&-5.4    			&  -5.7\\
max(background)                  			&4.0     			&    4.1\\
$<$line$>$\tablenotemark{b}                     &12.6   			&  29.6\\
$\sigma$(line)                       			&  5.6   			&  5.8\\
min(line)                               			&  -18.2 			&4.8\\
max(line)                               			&35.7   			& 54.9\\
$<$line$>$/$<$background$>$			&11.5			&26.9\\    
\\
\enddata
\tablenotetext{a}{Intensities in MJy~\psr}
\tablenotetext{b}{Sum over 5 channels as illustrated by Figure \ref{H2_17_spec_exp}}
\end{deluxetable}

We see a very small ($<$1 MJy\psr) residual background that remained after subtracting data from reference region from the two mosaics of the Taurus boundary region. 
This was essentially a constant function of wavelength and was removed by fitting a linear baseline to the spectrum of each pixel of the mosaics.
Pixels with line intensities greater than $\simeq$5 MJy\psr\ are statistically significant detections, which includes most of the positions in the Edge region and almost all of the positions in the Peak region.  
As indicated by the extreme values of maximum and minimum intensity, there are still are some defective pixels.  
But from the distributions shown in Figure  \ref{Inten_Stats}, it is clear that these are a very small fraction of the pixels in the images of the two regions.

The mean intensities correspond to column densities $N(J =3)$ = 2.3$\times$10$^{17}$ \cmc\ in the Edge region and 5.1$\times$10$^{17}$\cmc\ in the Peak region.
For comparison, the column density derived from {\it Spitzer} data in the Edge region was 1.0$\times$10$^{17}$ \cmc, and in the Peak region was 1.6$\times$10$^{17}$ \cmc \citep{Goldsmith08}
Thus, the present results are a factor 2--3 times greater.
The much larger extraction aperture used for {\it Spitzer} data reduction compared to size of JWST images as well as instrument calibrations are plausible contributors to this difference.

Beyond confirming the presence of highly excited \hh\ in the Taurus boundary initially detected by  {\it Spitzer}, the outstanding feature of the emission imaged by JWST in Figures \ref{fig_obs20} and \ref{fig_obs30} is the presence of small scale structures.
These show a variety of shapes, with sizes in the range between 1.0\as\ and 2.5\as, corresponding to physical dimensions 140 to 350 AU.
We discuss some of the characteristics of these clumps determined by the \hh\ emission below.
The sizes are consistent with results of analysis using power spectrum and other techniques; these are discussed in more detail in a paper currently in preparation. 
We feel the convincing detections in two regions do suggest that such small scale structures may be common throughout the molecular interstellar medium.
\section{Column Density of H$_2$}
\label{coldens}

\subsection{$J$ = 3 Column Density}

We assume that the H$_2$ rotational line emission is optically thin\footnote{For highly subthermal excitation and a line width of 1 \kms\ a column density $N$(ortho--\hh)$\simeq 10^{25}$ \cmc\ is required to reach an optical depth of unity in the $S(1)$ line, at least 5 orders of magnitude greater than we derive below.}.  
In this limit the column density in the upper level of the observed transition, $N_u$ (\cmc), is related to the observed intensity $I$ (erg \ps\cmc\psr) by
\beq
N_u = \frac{4 \pi I}{A_{u,l} h\nu} \lc
\eeq
where $A_{u,l}$ is the spontaneous decay rate(\ps), $h$ is Planck's constant, and $\nu$ is the frequency of the transition (Hz).
For the H$_2$ $J$ = 3$\rightarrow$1 S(1) line, the lowest transition of ortho--\hh, $A_{3,1}$ = 4.76$\times$ 10$^{-10}$ \ps 
and $\lambda$ = 17.035 $\mu$m \citep{Roueff19}, corresponding to $\nu$ = 1.76$\times$10$^{13}$ Hz.  
Although JWST/MIRI does not resolve the \hh\ emission from interstellar clouds, the pipeline output is given in units of specific intensity, $I_{\nu}$ (erg \ps\cmc\psr\pHz), with the relationship with the upper level column density thus being
\beq
N_u = \frac{4 \pi I_{\nu} \delta\nu}{A_{u,l} h\nu} \lc
\eeq
where $\delta \nu$ is the frequency resolution, taken to be the full width at half maximum of the spectral response curve.   
The ratio of the observed frequency $\nu$ to the frequency resolution $\delta \nu$ is the resolving power, $R$, so that the preceding equation can be written
\beq
N_u = \frac{4 \pi }{A_{u,l} h R} I_{\nu} \lp
\eeq
The resolving power of JWST/MIRI is wavelength dependent.  We here adopt the value $R$ = 2500, slightly above value 2443 of the fitted curve for spatially unresolved sources given by \citet{Jones23}, allowing for slight broadening due to the extended nature of the emission observed here.

The coefficient relating specific intensity to column density  is 2.58$\times$10$^{34}$ \cmc/erg \ps\cmc\psr\pHz\ for the S(1) line with the above parameters.
The specific intensity units of the JWST data output are MJy \psr.  
As 1 MJy \psr = 1.0$\times$10$^{-17}$ erg \ps\cmc\psr\pHz, the conversion we employ for the $J=3$ upper level of the $S$(1) line is

\beq
N(J=3) ({\rm cm}^{-2}) = N_u = 1.6\times 10^{16} I_{\nu} (S(1)~({\rm MJy~sr}^{-1}) \lp
\eeq
The intensity corresponding to a 1 MJy \psr\ $S$(1) line observed with JWST/MIRI is 7.04$\times$10$^{-8}$ erg \ps\cmc\psr.
%
%
\section{\hh\ Excitation and Total Column Density of $ortho$--\hh}
\label{extotal}

A number of different mechanisms have been proposed to populate the excited rotational levels of \hh\ and explain the observed far--infrared line emission \citep[e.g.][]{Falgarone05, Neufeld06, Goldsmith10}. 
These include excitation by cosmic rays, absorption of UV radiation, radiative decay of newly--formed \hh, and collisional excitation.  
We here briefly discuss each of these.
Figure 7 of \citet{Padovani22} shows a schematic of several of the various mechanisms.

The total column density of ortho--\hh\ relative to that in $J$ = 3 level is determined by the excitation of the molecule.
For \hh, only the lower rotational levels have significant population at even the highest temperatures considered.  
Since the spontaneous decay rates for vibration--rotation lines are ~100 to ~1000 times faster than those of the pure rotational transitions, the fractional populations of levels with $v \ge 1$ are negligible.  For example, even in LTE at 1000 K, the population of the $v$ = 1, $J$ = 1 level with $\Delta E/k$ = 6149 K is only 0.0025 of $v$ = 0, $J$ = 1, and drops precipitously at lower temperatures.

\subsection{Cosmic Ray Excitation}
\label{CRex}

Cosmic ray (CR) protons and electrons and the secondary electrons they produce can ionize \hh. Secondary electrons from CR protons and electrons can excite vibrational and electronic levels of \hh.
These processes have been studied in some detail \citep[e.g.][]{Gredel95}, with \citet{Padovani22} giving updates to rates and an overview discussion.  
The cosmic ray ionization rate, $\zeta_{cr}$  has been measured in a wide range of environments using different techniques.
In dense clouds, \citet{Caselli02} found that best--fit models had 5$\times$10$^{-19}$ $\le$ $\zeta_{cr}$ $\le$ 5.0$\times$10$^{-18}$ \ps, and \citet{Williams98} determined $\zeta_{cr}$ = 5$\times$10$^{-17}$ \ps in a sample of cores having densities 1--3$\times$10$^4$ \cmv.  
\citet{Bialy22} used ground--based observations of \hh\ rovibrational transitions to derive an upper limit $\zeta_{cr}$ $\le$ 1.5--3.6$\times$10$^{-16}$ \ps\ in the interiors of four clouds with column densities $N$(\hh) = 10$^{22}$ \cmc.
In less dense regions, the inferred cosmic ray ionization rate is considerably higher, ranging up to 4$\times$10$^{-16}$ \ps \citep{Vandertak06}, and even $\simeq$10$^{-15}$ \ps \citep{McCall03, Oka05}.

A useful quantity for the analysis of \hh\ excitation is the product of the spontaneous decay rate and the upper level volume density. 
Let us consider the ``clumps'' clearly seen in the image of the \hh\ $S(1)$ intensity distribution. 
We take the line of sight path length of a clump to be equal to a typical transverse size of 1\as =  2.1$\times$10$^{15}$ cm.  
With a characteristic value \Inu\ = 10 \MJysr, $N_u$ = 1.6$\times$10$^{17}$ cm, and $n_u$ = 76 \cmv.
This yields for the product
\beq
\label{nA}
n(3)A_{3,1}  = 3.6\times10^{-8} {\rm cm}^{-3} {\rm s}^{-1} \lp
\eeq

We carry out an approximate analysis by assuming that the excitation by secondary electrons from CR protons or electrons excites the \hh\ molecule from its low--$J$ state in the ground vibrational state to a vibrationally excited level with $v$ $\ge$ 1, but often higher.  
As mentioned above, the rovibrational transitions decay extremely rapidly compared to pure rotational transitions.  
Collisional deexcitation is less rapid than spontaneous decay given the densities in question, as discussed below.
After being excited, there is thus a series of rovibrational decays.  
In this process, the ortho-- and para--\hh\ remain separate species.  
The ortho--\hh\ could end up back in the $v$ =1, $J$ = 1 state (having decayed from $v$ = 1, $J$ = 1 or $v$ =1, $J$ = 3 state), but is more likely to have decayed to a higher--$J$ state in $v$ = 0, and to then cascade down via a series of pure rotational transitions.  

Without doing a complex multilevel calculation we can obtain an upper limit to the rate of the population of the $v$ = 0, $J$ = 3 level by assuming that every excitation of \hh\ results in an \hh\ molecule in the $J$ = 3 level.
In this model, the rate equation for the $J$ = 3 level in $v$ = 0 is
\beq
\label{crbal}
n(1)\zeta^e_{cr} = n(3)A_{3,1} \lc
\eeq
where $\zeta^e_{cr}$ is the cosmic ray excitation rate and we have assumed that all ortho--\hh\ is essentially all in the $J$ = 1 rotational level of the ground vibrational state due to the highly subthermal excitation.
The ratio of the excitation rate to the cosmic rate ionization rate is $\simeq$2 for \hh\ column densities $\leq$ 10$^{20}$ \cmc, but drops by $\leq$ 10\% for \hh\ column densities as large at 10$^{22}$ \cmc \citep{Padovani22}
\footnote {The calculations of \citet{Padovani22} are for \hh\ in the $J$ = 0 state as appropriate for very low $OPR$ but the rates should not be very different for \hh\ in $J$ = 1. }.

We model a clump seen in the \hh\ $S(1)$ emission with emission 5 MJy~\psr\ above the extended emission, and in consequence a column density $N(3)$ = 8$\times$10$^{16}$ \cmc.  
Adopting a characteristic size of 1\as, we obtain the volume density $n(3)$ = 38 \cmv, which yields $n(3)A_{3,1} \sim 1.8 \times$10$^{-8}$ \cmv \ps.
Even if we assume the extreme conditions $\zeta_{cr}$ $\simeq$1$\times$10$^{-15}$ \ps\ and thus $\zeta^e_{cr}$ $\simeq$ 2$\times$10$^{-15}$ \ps, with \hh\ $\sim$ 1000 \cmv, we obtain $n(1)\zeta^e_{cr} \ll n(3)A_{3,1}$. We conclude that cosmic ray excitation cannot be responsible for the observed \hh\ emission.

\subsection{UV Excitation}
\label{UVex}
A general category of processes that can produce \hh\ emission is the radiative cascade resulting from newly--formed \hh\ molecules, with one source of such molecules being UV radiation.
Modeling effect of UV on \hh\ is complex, as the interaction of the radiation with the molecules can take many forms.  
In addition to external UV radiation, there is an internal radiation field produced by cosmic rays that can excite, photoionize, and  photodissociate the \hh\ \citep[e.g.][]{LeBourlot95}, although these processes can also be included as part of the effects of cosmic rays.
\citet{Black87} considered the effects of UV in detail, but calculated only the emission in the near--infrared rovibrational transitions.
The UV can excite the \hh, which will result in a radiative cascade that populates the lower levels.
If the \hh\ is destroyed, it can reform, and in general a significant fraction of its binding energy goes to internal degrees of freedom, resulting in population of mid-- to high--lying vibrational levels.
These again cascade downwards and can produce emission in the pure rotational lines as well as the more often studied near--infrared rovibrational transitions.  
In some treatments of  \hh\ formation cascade, the process involved can be identified by the characteristic far--infrared emission spectrum \citep[e.g.][]{Tine03}.
In other models \citep{Islam10}, the newly--formed \hh\ is significantly vibrationally excited (3 $\leq$ $v$ $\leq$ 5) but rotationally cool ($J$ = 1 has by far the largest fractional population of ortho--\hh).  

\citet{Takahashi01} considered the effect of different types of grains on the distribution of formed \hh\ over different internal states, and included pure rotational as well as rovibrational transitions.
They modeled a cloud having total proton density of 10$^3$ \cmv, but which was evolving from atomic to molecular form, and thus almost all hydrogen would be H$^0$.  
For all models they considered, the intensity of the $S(1)$ line is 5.5--6.3$\times$10$^{-8}$ erg \ps \cmc \psr.  
This is only about 1\% of that corresponding to our detected intensity of 10 MJy \psr, indicating that even this situation with the maximum possible \hh\ formation rate cannot explain the observed intensity.

A general treatment showing the inadequacy of formation cascade population of the $J$ = 3 level can be shown using a similar treatment to that employed for analysis of cosmic ray excitation of \hh.  
In this case, we assume that the \hh\ formation rate is given by 
\beq
\frac{dn({\rm H}_2)}{dt} = k_f n_{\rm H} n({\rm H}^0) \lc
\eeq
where $k_f$ is the rate coefficient which is found to be 1.5--2.0$\times$10$^{-17}$ {\rm cm}$^3$\ps by \citet{BIaly24},  $n$(H$^0$) is the atomic hydrogen density, and $n_{\rm H}$ = $n$(H$^0$) +2$n$(H$_2$).
If we again assume that all newly--formed \hh\ ends up in the $v$ = 1, $J$ = 3 level, the rate equation for the $J$ = 3 level becomes
\beq
 k_f n_{\rm H} n({\rm H}^0) = n(3)A_{3,1} \lp
 \eeq
 In the limit that essentially all hydrogen is atomic, we obtain the maximum rate into the $J$ = 3 level, given by  $k_fn^2({\rm H}^0)$ \cmv\ps, and for $n$(H$^0$) = 10$^3$ \cmv, this rate is 2$\times$10$^{-13}$ \cmv\ps.
 This is 5 orders of magnitude less than the rate of transitions from $J$ = 3 to $J$ = 1, and confirms the inadequacy of formation cascade to explain the observations.

UV excitation (not involving \hh\ destruction) can be more rapid due to trapping and effective reuse of photons, but the increase in the rate of populating the $J$ = 3 level is still certainly less than required.
If the \hh\ were in chemical steady--state and destruction were dominated by UV, then the formation rate would be equal to the UV photodestruction rate.  
However, the UV radiation field in the boundary regions of Taurus is relatively low \citep{Flagey09}, 0.3 -- 2.8 $\times$ the standard Habing radiation field (1.6$\times$10$^{-3}$ erg\ps\cmc), and the resulting destruction rate would result in an even lower \hh\ formation rate and cascade population of the $J$ = 1 level than modeled above.

\subsection{Collisional Excitation}
\label{COLLex}

Collisional excitation of \hh\ rotational levels will certainly occur, with rate depending on temperature and density.  
The rate coefficients for collisions with \hh\ calculated by \citet{Flower98} depend only very weakly on the spin modification of the colliding partner, which is assumed to be in $J$ = 0 for para--\hh\ and in $J$ = 1 for ortho--\hh.
The gas--phase hydrogen in these regions of interest may not be entirely molecular, and in this case collisions with hydrogen atoms must be considered.  
The deexcitation rate coefficients calculated by \citet{Forrey97} are approximately a factor of 3 smaller than those for collisional deexcitation by \hh.
The molecular hydrogen fraction in the regions observed is not known, and for simplicity we assume that the hydrogen is completely molecular. 
At a temperature of 300 K, assuming that the translational velocity distribution of the \hh\ molecules is thermal, the critical density for the $S(1)$ line is 366 \cmv, so that the excitation temperature of the transition is likely neither highly subthermal nor thermalized.

\subsubsection{Ortho to Para Ratio}
\label{OPRrat}

The total \hh\ column density depends on the ortho-to-para ($OPR$) ratio of the \hh, with the relationship between the total \hh\ column density $N$(\hh), the total column density of ortho--\hh, $N(ortho)$, and the $OPR$ being
\beq
N(\hh) = N({\rm ortho})(1 + OPR^{-1}) \lp
\eeq
In equilibrium at low temperatures, the $OPR$ will be very small, while \hh\ is formed with eV of energy, and thus is effectively hot, with $OPR$ =  3, reflecting the ratio of statistical weights of the two spin modifications.  
Our data of a single transition cannot determine the $OPR$ and observations of the $OPR$ in quiescent \hh\ are difficult and rare.  
Previous {\it Spitzer} observations of \hh\ in this same region yielded $OPR$ = 0.95 \citep{Goldsmith10}.

The time-dependent evolution of the $OPR$ in a region with changing temperature is complex, as conversion between the two spin modifications can involve gas phase collisions with H atoms and ions as well as grain surface reactions \citep{Lique12, Bron16}.  
If conditions in the very small regions being considered here have been evolving, the $OPR$ will not simply reflect the temperature, but will depend on the past history of the gas.  
This likely starts with an extremely low $OPR$ at the temperature of the molecular cloud boundary, and increasing when the temperature increases.  

The conversion rate of para-- to ortho--\hh\  conversion by exchange in collisions with H atoms is very slow, $\le$10$^{-15}$ cm$^3$s$^{-1}$ for $T$ $\le$ 400 K \citep{Lique12}.  
This yields a characteristic time scale $\sim$ 10$^5$ yr at a density of 100 \cmv, which is significantly longer than the characteristic timescales of TDRs or shocks (discussed below).
Thus, the $OPR$ is not easily determined by modeling.

Observations of four star--forming regions producing shocked \hh\  \citep{Neufeld06} suggested two values of the $OPR$:  0.2--0.5 and 1.5--2 for the warm and hot component, respectively. 
Numerous studies have found that \hh\ in photon dominated regions (PDRs), $OPR$ $\sim$ 1 despite gas temperatures of several hundred to one thousand K \citep[see references on p.2 of][]{Bron16}.
Given the available information we adopt $OPR$ = 1, so that $N$(\hh) = 2$N$(ortho--\hh).
Since the $OPR$ is unlikely to be less than unity (a value characteristic of equilibrium at typical diffuse ISM temperatures) and is restricted to be less than 3 (except in very unusual cases) for which $N$(\hh) = 1.33$N$(ortho--\hh),  the uncertainty introduced by our lack of precise information on the $OPR$ is modest.

\subsubsection{Total \hh\ Density and Kinetic Temperature}

\begin{figure*}[htb!]
\center
\includegraphics [angle=0, width=12 cm]{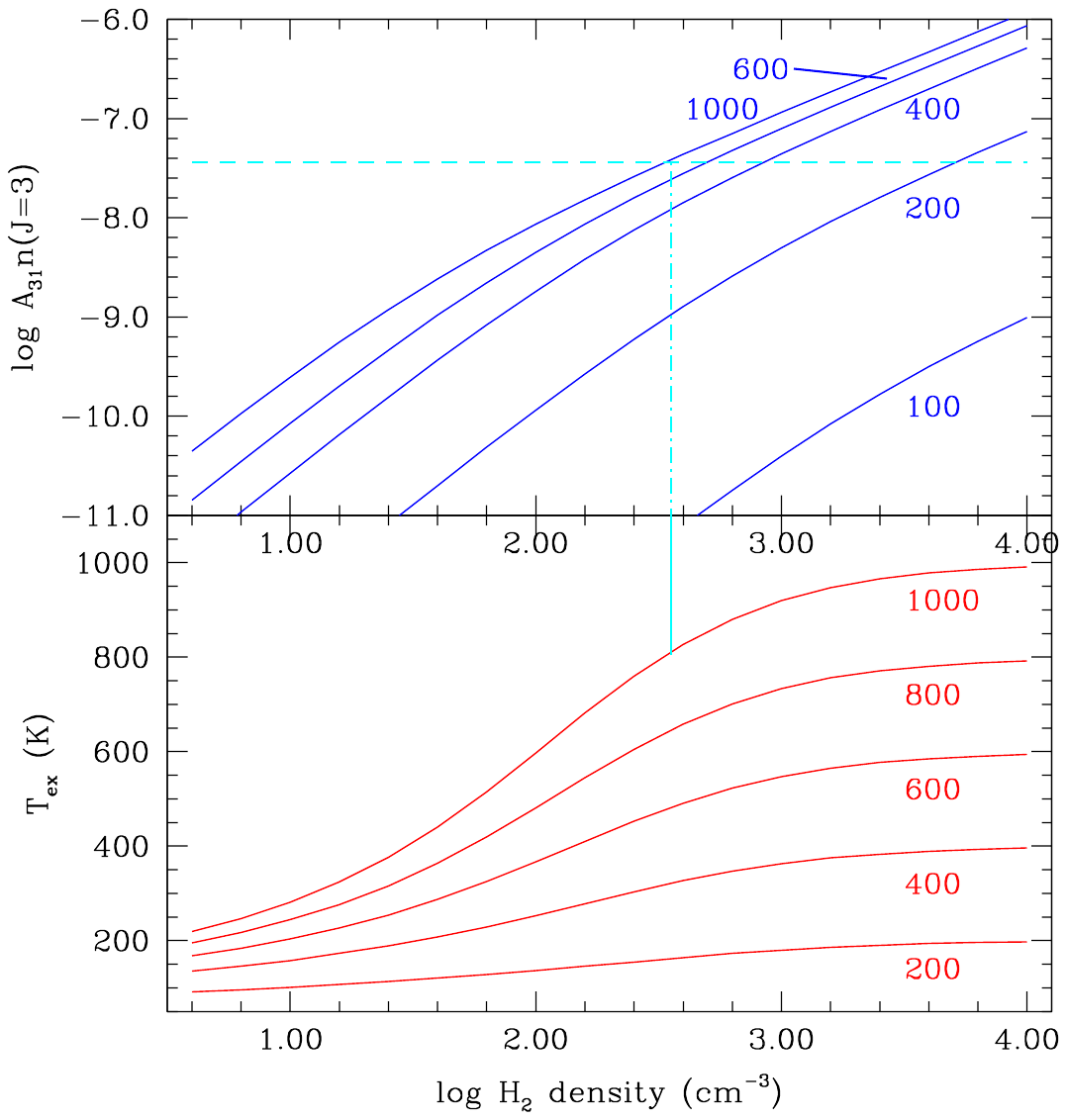}
\caption{\label{nATex} Upper panel:  Dependence of the quantity $n(3)A_{3,1}$ characterizing the emission from the J = 3 level of ortho--\hh\ on total \hh\ density and kinetic temperature.  
The five blue curves are labeled by their kinetic temperature.  The horizontal dashed cyan line shows value, log($n(3)A_{3,1}$) = -7.44, corresponding to an observed intensity of 10 MJy\psr. 
The intersection of this line with curve for given kinetic temperature (here 1000 K) determines the total \hh\ density (here $\simeq$350 \cmv), as indicated by the vertical dot--dashed cyan line.
Lower panel:  excitation temperature of the $S(1)$ line as function of total \hh\ density and kinetic temperature. 
The vertical solid cyan line indicates the excitation temperature, 800 K, determined by the selected kinetic temperature and the derived \hh\ density. }
\end{figure*}

Solving the rate equation for lowest levels, we can calculate the quantity $n(3)A_{3,1}$ (Equation \ref{nA}) as a function of the total \hh\ density and kinetic temperature (T$_{kin}$).  
The result is shown in the upper panel of Figure \ref{nATex}.  
The lower panel shows the excitation temperature of the $S(1)$ line.
From a given value of $n(3)A_{3,1}$ we find the required total \hh\ density for a given kinetic temperature, as well as the excitation temperature characterizing the degree of excitation of the \hh.

It is evident that a lower kinetic temperature requires a higher total \hh\ density to produce the observed emission, and that minimum kinetic temperature of $\ge $ 200 K is required to have a density $\le$ 10$^4$ \cmv.

In this model, the thermal pressure of the small structures seen in the JWST images is much better constrained than the temperature and density individually.
If we exclude the lowest kinetic temperature, we find that for 400 K $\le$ T$_k$ $\le$ 1000 K the pressure, $P/k$ is within 10\% of 3.5$\times$10$^5$ K \cmv.
This is far greater than that of the CNM,
with $<P/k>$ = 3800 K \cmv\  \citep{Jenkins11}; 2000 K \cmv\ $\leq$ $P/k$ $\leq$ 10,500 K \cmv\ \citep{Goldsmith18}.  It is also far above the pressure measured in diffuse molecular gas, for which $P/k$ is between 4600 and 6800 K \cmv\ \citep{Goldsmith13}.
The thermal pressure is comparable to that found in TSAS observed in the radio, with $P$ $\simeq$ 10$^6$ K \cmv\ \citep{Stanimirovic18}.  
The size of these very small regions, 1 $\leq$ $L$ $\leq$ 100 AU is comparable to that of the structures we have found in the \hh\ $S$(1) emission.
The extreme overpressure of the \hh\ condensations suggests that their lifetime is limited, but this would be a natural result of their being turbulent dissipation regions or postshock instabilities (discussed below).

\begin{figure*}
\center
\includegraphics [angle=0, width=12 cm]{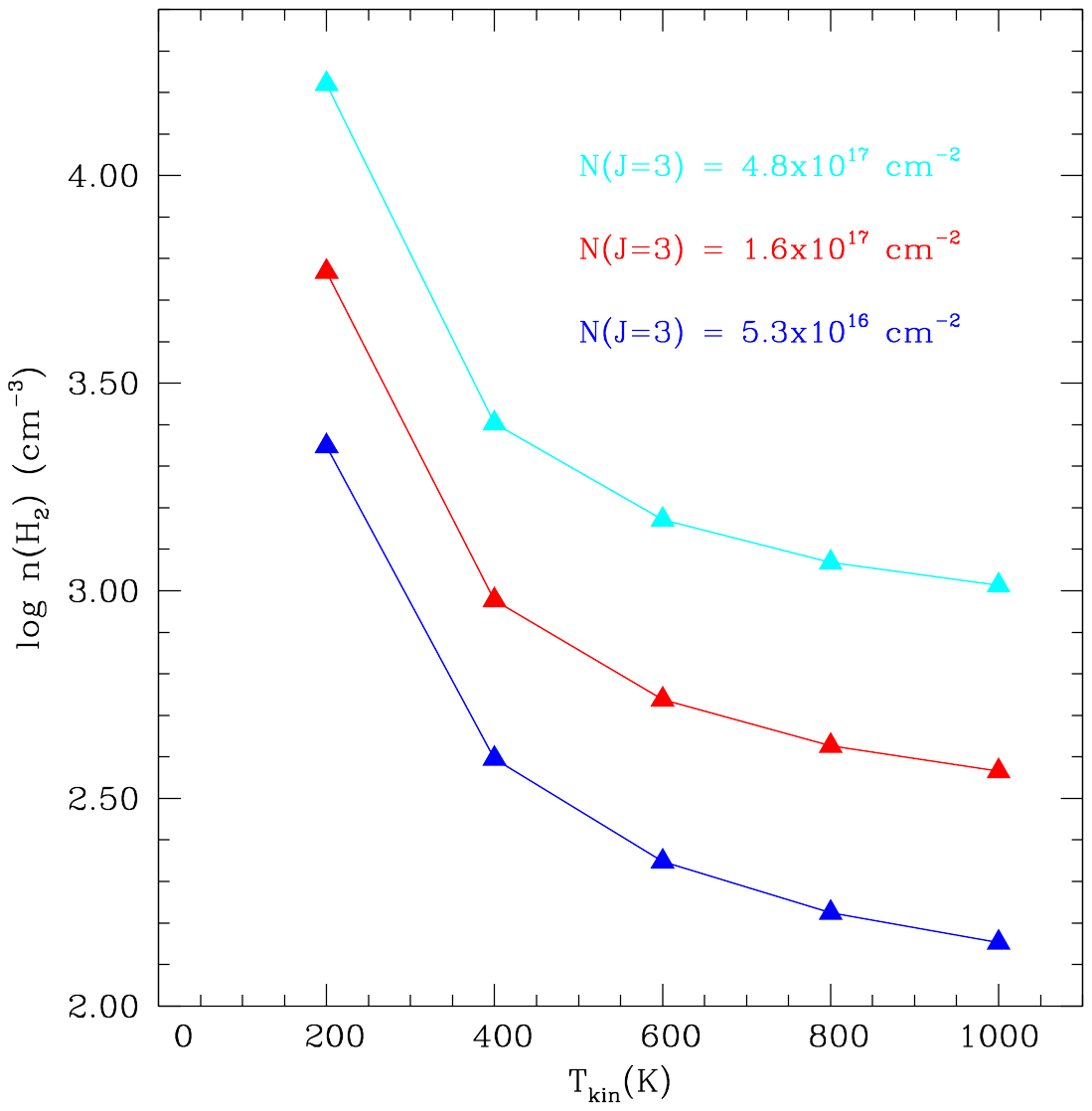}
\caption{\label{quad} Total \hh\ density as function of kinetic temperature for  given $J$ = 3 column density, $N(J = 3)$.  The three curves are for the column density corresponding to intensity 10 MJy\psr, and factor of 3 greater and smaller column density than this.}
\end{figure*}

We can simplify the problem by considering only the two lowest levels of ortho--\hh\ in the excitation calculation.  
This is justified by the much higher energy of the $J$ = 5 level, $E/k = 2333$ K compared to that of $J$ =1, and the much faster spontaneous decay rate, $A_{5,3}$ = 9.84$\times$10$^{-9}$ \ps\ $\simeq$20 times greater than $A_{3,1}$.  
The result is that there is negligible population in $J$ = 5, as any infrequent collisional excitation of that level decays rapidly to $J$ = 3.
With the definition $f(3)$ as the fraction of ortho--\hh\ in the $J$ = 3 level we can write for the total \hh\ density
\beq
\label{nh2tot}
n(\hh) = \frac{n(3)(1 + OPR^{-1})}{f(3)} \lp
\eeq
The collision rate is the product of the collision rate coefficient and the total \hh\ density since either ortho-- or para-- \hh\ can excite the $S(1)$ transition
\footnote{In Appendix B we determine that the total extinction through the Taurus Molecular Cloud boundary region is 0.2 $\le A_v\le $ 1.0 mag, so that the hydrogen will be largely molecular.  
We also consider only moderate temperatures in {\it e.g.} postshock regions, which are too low to destroy a significant fraction of the \hh. 
Thus, assuming the excitation is by collisions with \hh\ is reasonable.
}.
The upwards collision rate is $C_{1,3}$ = $R_{1,3}n$(\hh) and  the downwards collision rate  $C_{3,1}$ = $R_{3,1}n$(\hh), with $R_{1,3}$ = $R_{3,1}(7/3)exp[-(845/T_{kin})]$.  
The rate equation gives
\beq
\label{f3}
f(3) = \frac{C_{1,3} + C_{3,1} + A_{3,1}}{C_{1,3}} \lp
\eeq
Inserting Equation \ref{f3} into Equation \ref{nh2tot} yields a quadratic equation for $n$(\hh), with a solution in the usual form 
$n({\rm H}_2) =( -b + (b^2 - 4ac)^{0.5})/2a$,  where a = $R_{1,3}$, b = $-(1 + OPR^{-1})n(3)(R_{1,3} + R_{3,1})$, and c = $-(1 + OPR^{-1})n(3)A_{3,1}$.
For a given $S(1)$ intensity, we have the column density in $J$ = 3, $N(3)$, and with the assumed size of the region $L$, we get the observed density in $J$ = 3, $n(3)$.  The solution is the total \hh\ density as a function of the kinetic temperature.
Figure \ref{quad} shows results for the nominal column density $N(3)$ = 1.7$\times$10$^{17}$ \cmc, and a factor of 3 higher and lower column density.

While lower kinetic temperatures cannot be {\it a priori} excluded, a total \hh\ density $\le$500 \cmv\ required for $T_{kin}$ $\ge$ 600 K appears more reasonable than the much higher densities required for lower kinetic temperatures.  
The total \hh\ column density required for $T_{kin}$ = 200 K is approximately an order of magnitude larger than that required for $T_{kin}$ = 600 K.  
Kinetic temperatures $\simeq$1000 K are far higher than those found in the quiescent molecular interstellar medium, and appear to be the only explanation for presence of certain ions, particularly CH$^+$, for which the key formation reaction is highly endothermic \citep[e.g.][]{Falgarone10a, Falgarone10b}.
This is clearly vastly hotter than the 10 K $\le T_{kin}\le$ 100 K found in the molecular ISM \citep{Goldsmith87} and also the cold neutral medium \citep{Heiles03a}.
Several mechanisms have been proposed to explain such elevated temperatures in small regions of the ISM, and we briefly discuss these in the following section.


\section{Producing the High Kinetic Temperature Required to Explain the \hh\ $S(1)$ Emission}
\label{highT}

We here discuss two possible sources for localized heating of the interstellar medium that could explain the structure in the \hh\ emission that we have observed with JWST; these are dissipation of turbulence and shocks.

\subsection{Dissipation of Turbulence and Small-Scale Heating}
Turbulence is well-recognized as a general characteristic of the interstellar medium. 
As discussed in the review by \citet{Falceta14}, it is characterized by fluctuations in time and space of various quantities including velocity and density.
The general picture is that energy is injected on a relatively large scale, and cascades down to a much smaller scale of tenths of AU on which there is dissipation due to viscosity, which produces local gas heating \citep{Falgarone95}. 

This led to specific models for turbulent dissipation involving magnetized vortices \citep{Falgarone95b, Joulain98, Godard09, Godard14}.
These models have been tailored for a largely atomic medium, but the physical characteristics are in many respects very similar to what we have derived for the small--scale structures seen in the Edge and Peak regions of the Taurus Molecular Cloud boundary.
Figure 2 of \citep{Godard09} indicates a peak temperature of $\simeq$700 K at a radius of 50 AU for their reference model, having $n_{\rm{H}}$ = 30 \cmv and $A_v$ = 0.4 mag.  
This density is a factor $\simeq$10 lower than the minimum we can find acceptable to explain the density of \hh\ in $J$ = 3 state.
The picture of dissipation in these models is that there are many such small turbulent dissipation regions (TDRs) along the line of sight, so that each has much weaker emission than we observe.
However, the range of density and size studied by \citet{Godard14} comes close to including parameters that would explain our \hh\ results.

\citet{Lesaffre20} carried out a very high spatial resolution modeling of localized heating resulting from dissipation of turbulence.
A random velocity field with $rms$ velocity 2.1 \kms\ in a region with $n_{\rm H}$ = 100 \cmv\ and $N_{\rm H}$ = 1$\times$10$^{18}$ \cmc\ led to turbulent dissipation and to the formation of shocks, which raised the temperature to over 1000 K.
However, the excitation temperature of the $S(1)$ line ($\simeq$150 K) is considerably less than we estimate (Figure \ref{nATex}) because the lower density does not provide the required collision rate.
Further high--resolution numerical modeling is necessary to understand whether turbulent dissipation operating on the scale of structures we observe and in a largely molecular medium can provide the observed \hh\ emission.

\subsection{Shocks}
\label{shocks}
Shocks in the interstellar medium perturb the velocity of the gas, increase its density, generally producing also an increase in temperature.  
There are many different types of shocks, with and without magnetic fields, which differently affect the gas, and various models treat these in different ways, including both individual and multiple shocks being present.
Without being comprehensive we here give some of the results that are relevant for \hh\ emission.
As illustrated by the spectra of \hh\ shown by \citet{Neufeld06}, shocks can broaden line profiles, and if the shock velocity is sufficiently large, the presence of shocks can be almost unambiguously identified.  
Since we are observing regions without significant sources of mechanical energy ({\it e.g.} young or evolved stars), we expect any shock velocities to be less than the 120 \kms\ resolution of  JWST MIRI/MRS at 17 $\mu m$ wavelength.

We present \co\ spectra in the directions of the Peak and Edge regions in Appendix B.  
The line parameters indicate that the hydrogen along these lines of sight is largely molecular.
The line profiles of \co\ in both regions clearly suggest that multiple velocity components are present; these are also suggested in the \coo\ spectra with marginal signal to noise ratio. 
In both positions, the velocities of the various components  differ by $\simeq$ 3 \kms.
It is possible that these represent different gas components that are colliding to produce low--velocity shocks.
While much higher--angular resolution data would be valuable, we consider the possibility of low--velocity shocks being responsible for the heating of the gas that is responsible for the \hh\ emission.

\citet{Pon12} considered few \kms\ shocks propagating in molecular gas with density of 1000 \cmv.  
These authors found that the \hh\ $S(1)$ intensity could be 5 to 15 times stronger than that found from the photon dominated region (PDR) associated with the molecular cloud, and thus potentially a discriminant between shocks and PDR emission.
Subsequent observations  \citep {Pon15} of mid--$J$ CO lines, particularly the $J = 9 \rightarrow 8$ transition, were interpreted as confirmation of the presence of such shocks.
However, the intensities predicted by three models presented in Table 4 of  \citet{Pon12} are $\sim$2 orders of magnitude below the Taurus boundary results, so this particular model cannot be explaining our observations.

\citet{Kristensen23} considered a very large grid of shock models, with different values of  initial density,shock velocity, $u$, and magnetic field parameter $b$  = $B_{\rm transverse}/n({\rm H})^{0.5}$.  
The $S(1)$ emission is maximized for the least magnetic shocks ($b$ = 0.1), and even for $u$ = 5 \kms\ is over an order of magnitude weaker than observed, although the peak temperature and density are significantly higher than required to be consistent with results.
Optimization of parameters would plausibly improve the agreement, but the gap between this model and our observations appears significant.

\citet{Lesaffre13} considered a wide range of shocks of different characteristics, with velocities between 3 \kms\ and 40 \kms, propagating in media with density between 10$^2$ and 10$^4$ \cmv.  
These authors find for weakly magnetized shocks in region in which hydrogen is molecular that the maximum temperature is given by 

\beq
T_{max} = 53~{\rm K} (u/ 1 \kms )^2 \lp
\eeq
Thus, a shock velocity of $\sim$ 3 \kms\ produces a maximum temperature $\sim$500 K.
As shown in Figure 1b of \citet{Lesaffre13}, the heated layer in such a low--velocity shock  in region with preshock density $n_{\rm H}$ = 100 \cmv\ has a thickness $\sim$10$^{15}$cm, comparable to the size of the small--scale structures we observe in the $S(1)$ emission.
The total \hh\ column density is 1$\times$10$^{18}$ \cmc, very close to what is implied by our results for kinetic temperaure = 500 K, $N$(\hh)  1.4$\times$10$^{18}$ \cmc. 
The intensity of $S(1)$ emission for a 3 \kms\ shock in their model is 1.4 MJy~\psr, significantly less than we observe.
For $u$ = 5 \kms, $I_{\nu}$ = 7 MJy~\psr, approaching the observed value, with temperature somewhat higher, and heated layer thickness smaller.
What is not clear is how the density throughout the extended heated layer compares to that we derive, namely $n$(\hh) $\simeq$ 700 \cmv\ for $T_{kin}$ = 500 K and $n$(\hh) $\simeq$ 400 \cmv\ for $T_{kin}$ = 800 K.
More detailed calculations including post--shock instabilities and formation of small-scale structures are necessary to more firmly connect the picture of low velocity shocks and our \hh\ observations.

\section{Summary and Conclusions}
\label{summary}
We have reported imaging of the $S(1)$ 17 $\mu m$ line of \hh with the MIRI/MRS instrument on the JWST.
Two regions on the boundary of the Taurus Molecular Cloud having total visual extinction $A_{\rm V}$ = 0.2--1.0 mag were observed using 4x4 mosaics of pointings to obtain, after some edge cropping, 69 x 81 0.2\as\ pixels.
We see \hh\ line emission at high statistical significance throughout most of the two regions, denoted ``Edge'' (near the  relatively sharp boundary of the \coo\ $J = 1\rightarrow 0$ emission) and ``Peak'' \citep[including the location of the strongest \hh\ emission observed with {\it Spitzer};][]{Goldsmith10}.  
The average flux densities for the Peak region and the Edge region are 32.1 MJy~\psr\ and 14.5 MJy~\psr\, respectively, corresponding to $J$ = 3 \hh\ column densities of 5.1 and 2.3 $\times$10$^{17}$ \cmc.

In addition to the extended emission, there are clearly isolated regions of stronger emission, having angular sizes 1.0\as -- 2.5\as, corresponding to physical sizes of 140--350 AU.
This is comparable to the tiny scale atomic structure seen in observations at radio wavelengths, but here evident for the first time in the quiescent molecular interstellar medium.
A characteristic flux density for these ``clumps'' is 10 MJy~\psr, corresponding to a column density of \hh\ in the $J$=3 level equal to 1.6$\times$10$^{17}$ \cmc.

We have investigated a number of different mechanisms for explaining this emission, which arises from the $J$ = 3 level 845 K above the $J$ = 1 level (the ground state of ortho--\hh) and 1016 K above the $J$ = 0 level (the ground state of para--\hh).
Cosmic ray excitation, UV excitation, and \hh\ formation cascade excitation all appear inadequate to explain the result in the relatively benign environment of the Taurus Molecular Cloud boundary.

The most plausible mechanism appears to be collisional excitation, dominated by collisions with other \hh\ molecules.
Assuming a characteristic length scale for the clumps of 1\as\ (2.1$\times$10$^{15}$ cm), we determine a set of solutions with given $n$(\hh) and T$_{kin}$ that can reproduce the observed 76 \cmv\ density of \hh\ in the $J$= 3 state.  
Higher temperatures are associated with lower densities, and while we cannot exclude, for example $n$(\hh) = 5$\times$10$^3$ \cmv\ with T$_{kin}$ = 200 K, considerably higher temperatures and lower densities are much naturally produced by two mechanisms that may be responsible for heating the clumps, namely turbulent dissipation and shocks.

A very reasonable solution is $n$(\hh) = 370 \cmv\ and T$_{kin}$ = 1000 K.  
This temperature can be produced by models of both turbulent dissipation and shocks.  
Models of turbulent dissipation regions, having been developed in part to explain the abundance of ions such as CH$^+$ which require a temperature close to this value to overcome activation barriers and endothermicity of key reactions, produce temperatures on the order of 1000 K.  
The peak postshock temmperature depends on the shock velocity, but modest shock velocities of 3 to 5 \kms\ are also capable of producing such temperatures. 

Shock models differ significantly in their predictions for the intensity of the $S(1)$ line, with that of \citet{Lesaffre13} coming reasonably close to the 10 MJy~\psr\ level required, for a weakly magnetized shock ($b$ = 0.1) with preshock density 100 \cmv\ and shock velocity 3 -- 5 \kms.  
The initial conditions of the modeled shocks typically are atomic gas, but the shock velocity required here is moderate (and so would not produce significant molecular dissociation) and is similar to that suggested by the extent of the components seen in the \co\ line profile.
We hope that future calculations will succeed in both showing the development of small--scale structure in the molecular interstellar medium and explaining the accompanying intense \hh\ $S(1)$ line emission that we have observed.  
JWST MRS studies of the $S$(2) and $S$(3) lines will help better constrain the properties, and so the origin, of this hot gas embedded in a quiescent interstellar medium.

\newpage

We thank Dr. Shmuel Bialy for insightful discussions about \hh\ excitation, Dr. Edith Falgarone for valuable information about TDRs, and Dr. David Neufeld for sharing his code for calculating molecular cooling in the type of region we have observed.
This research was conducted in part at the Jet Propulsion Laboratory, which is operated by the California Institute of Technology under contract with the National Aeronautics and Space Administration (NASA). Copyright 2025 California Institute of Technology.
This research was supported by the Space Telescope Science Institute through funding for JWST Cycle 2 Program GO-03050.  
X. W. is supported by the National Natural Science Foundation of China (grant 12373009), the CAS Project for Young Scientists in Basic Research Grant No. YSBR-062, the Fundamental Research Funds for the Central Universities, the Xiaomi Young Talents Program, and the science research grant from the China Manned Space Project.
X. W. also acknowledges work carried out, in part, at the Swinburne University of Technology, sponsored by the ACAMAR visiting fellowship. G.A.F gratefully acknowledges the Deutsche Forschungsgemeinschaft (DFG) for funding through SFB 1601 ``Habitats of massive stars across cosmic time'' (sub-project B1)  and from the University of Cologne and its Global Faculty programme. 
Support for R.S. was provided by NASA through the NASA Hubble Fellowship grant \#~HST-HF2-51566.001 awarded by the Space Telescope Science Institute, which is operated by the Association of Universities for Research in Astronomy, Inc., for NASA, under contract NAS5-26555.
\bibliography{bibdata}

\appendix

\section{JWST Data Processing Details}

We here provide a concise overview of the overall processing steps applied to the raw observational data and an introduction to the resulting dataset. 
The raw MIRI/MRS data were reduced using version 1.16.1 of the STScI pipeline and the CRDS context \texttt{jwst\_1303.pmap}. 
It is important to note that data reduced with pipeline versions earlier than 1.15.1 suffer from highly inaccurate error estimates. 

In the data reduction process, several non-standard settings were employed, including enabling cosmic ray shower flagging\footnote{\href{https://jwst-pipeline.readthedocs.io/en/stable/jwst/jump/description.html}{JWST pipeline cosmic ray snowball/shower information}}, applying 2D residual fringe correction\footnote{\href{https://jwst-pipeline.readthedocs.io/en/latest/jwst/residual_fringe/main.html}{JWST pipeline fringing information}}, and utilizing two different background subtraction algorithms (master background subtraction and pixel-by-pixel subtraction)\footnote{\href{https://jwst-pipeline.readthedocs.io/en/latest/jwst/background_subtraction/main.html}{JWST Pipeline  background subtraction information}}. 
Since our primary scientific goals focus on the S(1) and S(2) emission lines, no contamination from hot/warm pixels was detected at the corresponding wavelengths. 
Thus, manual masking of hot/warm pixels was not performed. However, during subsequent line identification, minor contamination due to hot/warm pixels at other wavelengths was identified. 
Addressing these contaminants lies outside the scope of this work.

One of the most critical challenges in data processing was reconciling the 4$\times$4 mosaicing design for S(1) with the background subtraction algorithms. 
Our target is an extended source that spans the observed field, requiring the spatial distribution of its emission lines to achieve our primary scientific objectives. 
For this purpose, pixel-by-pixel background subtraction was essential to ensure accurate flat-field correction. 
However, for spectral line identification, results from single exposures with master background subtraction were more suitable. 
Therefore, we produced datasets processed using both background subtraction algorithms and results without mosaicing for completeness.

\begin{figure*}[htb!]
\center
\includegraphics[angle= -90, width=12cm]{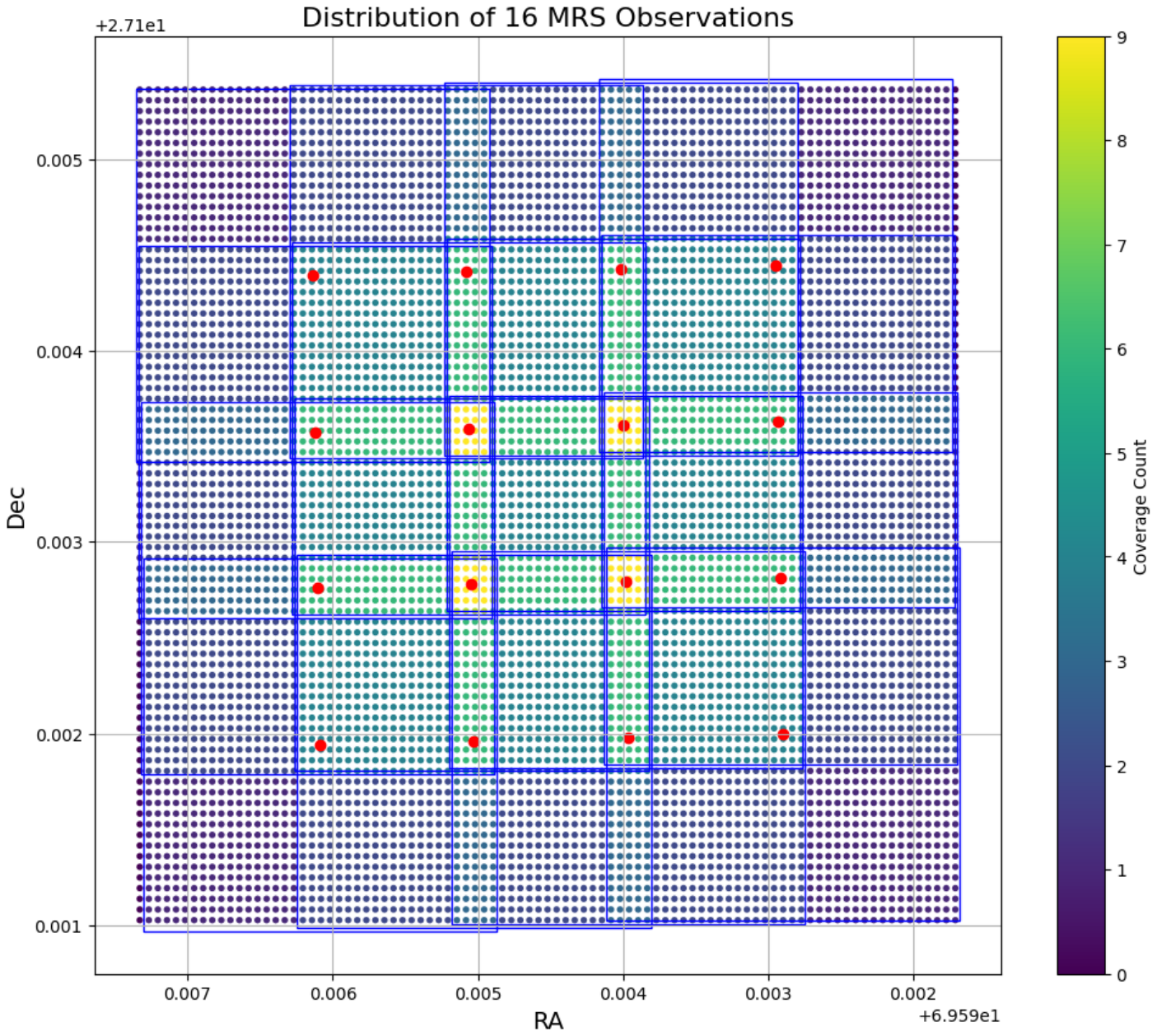}
\caption{\label{IntTime} Relative integration times for pixels in the 4 x 4 mosaic of pointings of Sub-band C, Channel 3 MIRI/MRS array used for observations of 17.04 $\mu \text{m}$ $S(1)$ line in the Taurus Boundary regions
}
\end{figure*}

    To ensure good overlap between adjacent single exposures, it is generally recommended to have a 10\% overlap between adjacent pointings in a 4-point dither pattern (in practice, the overlap after a 4-point dither is about 30\% in Channel 1).
    However, it is important to note that the FOVs of the four channels increase with wavelength. 
    To ensure good overlap across all four channels, we have used the FOV of Channel 1 as the reference. 
    This results in a significant increase in overlap from Channel 2 to Channel 4, with Channel 3 having an overlap of about 58\%. Considering the 4$\times$4 mosaic observation design, this leads to an interesting phenomenon—our observation depth is not uniform (see Figure \ref{IntTime}, and the distribution of the SNR generally decreases from the center to the outer edges. 
    When analyzing the spatial distribution of the data, we must account for the unevenness of the SNR.
Given that our focus is on the spatial structure of the molecular clouds, estimating the PSF scale is essential. 
However, no point sources were available in our observations for PSF modeling, and we consider the 4-point dither and mosaicing processes to have negligible effects on the PSF size. 
Therefore, we estimated the full width half maximum (FWHM) of the telescope PSF in the MRS wavelength range using the relationship provided by \citet{Law23},
\beq
\label{eq:PSF}
    \theta_{\rm FWHM}(\as) = 0.033 \, \lambda(\mu m) + 0.106.
\eeq
The PSF size directly constrains the lower limit of the molecular cloud structures that can be resolved in our observations. For S(1), this scale is $0.667 \arcsec$.
Our 4-point dither pattern allows us to Nyquist sample the FWHM of the PSF.

\section{CO Spectra}

We do not have velocity--resolved molecular data with the angular resolution of JWST or even of the size of our mosaiced observations.  
We do have \co\ and \coo\ $J = 1 \rightarrow 0$ data obtained with 45\as\ -- 50 \as\ resolution obtained using the FCRAO telescope \citep{Goldsmith08}. 
In order to achieve sufficient signal to noise ratio to search for any features due to low--velocity shocks, we averaged data over a region $\sim$8\am\ in size centered on the Edge and Peak positions; the resulting spectra are shown in Figure \ref{COspec}.
The line profiles do not appear to change significantly over the regions averaged to produce these spectra and the intensity exhibits only minor variations within each region.

\begin{figure*}[htb!]
\center
\includegraphics [angle=0, width=12 cm]{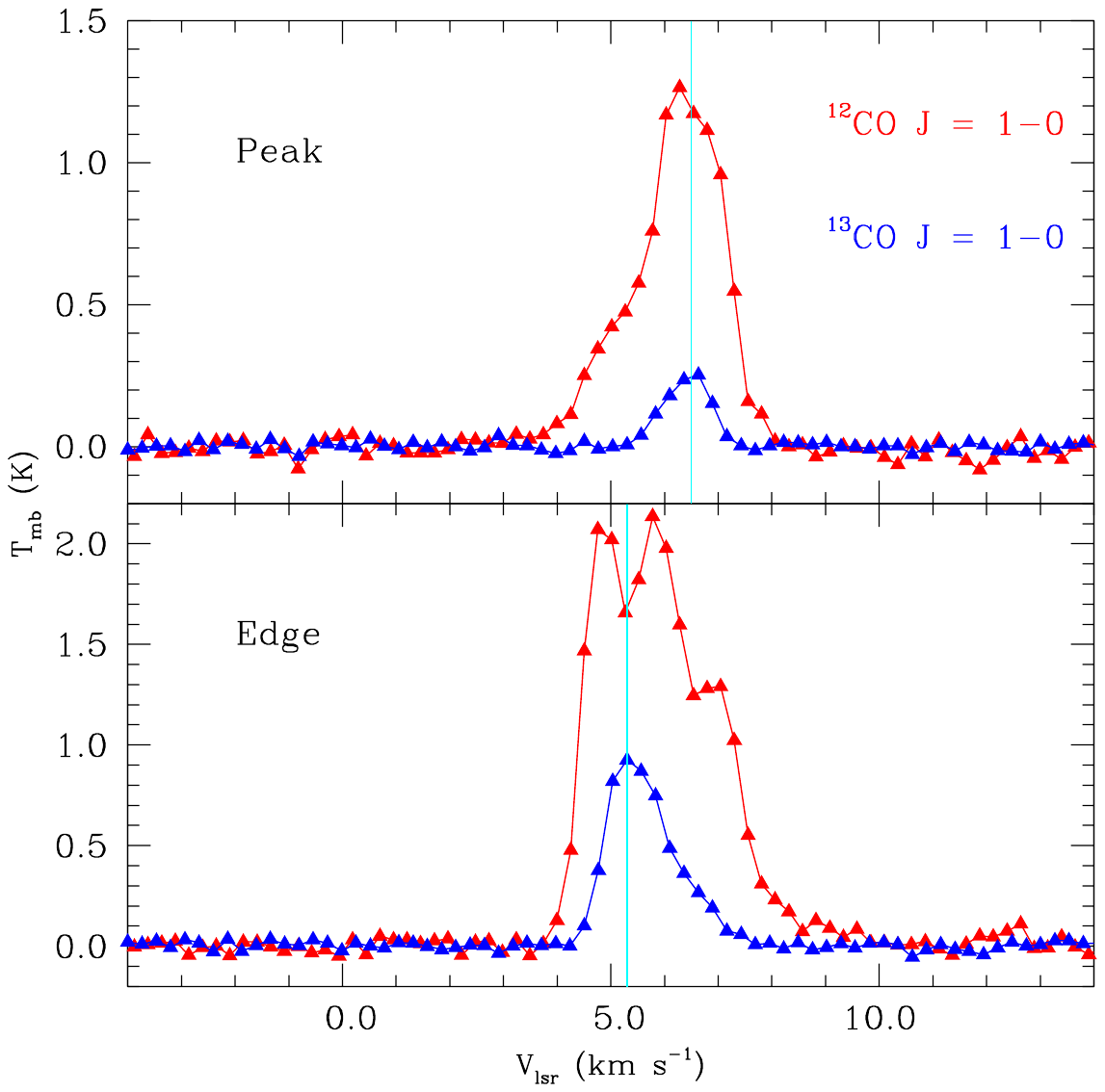}
\caption{\label{COspec} \co\ and \coo\ $J = 1 \rightarrow 0$ spectra from survey carried out with 14m FCRAO antenna \citep{Goldsmith08}.  Upper panl:  in region encompassing the Peak position.  Lower panel:  the same lines for region encompassing the Edge position.  
The vertical cyan lines identify the velocity of the strongest \coo\ emission at each position. \label{COspec}}
\end{figure*} 
The carbon monoxide emission seen in Figure \ref{COspec} likely does not come from the heated regions seen in the \hh\ $S(1)$ line.

Using RADEX \citep{Vandertak07}, we determine  $N$(\coo) = 6.0$\times$10$^{14}$ \cmc\ for the Peak and $N$(\coo) = 2.4$\times$10$^{15}$ \cmc\ for the Edge, assuming $\delta v$ = 1 \kms\ FWHM, $n$(\hh) = 100 \cmv, and $T_{kin}$ = 20 K.  
Assuming $N$(\co)/$N$(\coo) = 60, we find peak \co\ line temperatures approximately a factor of 2 greater than observed.
To bring the observed and modeled \co\ peak temperatures into agreement would require very low  $N$(\co)/$N$(\coo) ratio $\simeq$22, which could be a result of chemical isotopic fractionation \citep{langer80,langer84,furuya18}.  
It may also reflect variations in temperature and/or density along the line of sight, a filling factor less than unity, or self--absorption, as strongly suggested by the line profile of \co\ in the Edge spectrum.  
The \coo\ column densities suggest  \hh\ column densities in the range 2--10 $\times$10$^{20}$ \cmc\ and thus visual extinctions 0.2--1 mag.  
This is sufficient for the hydrogen to be largely molecular.

\end{document}